\documentclass[aps, prl, showpacs,  twocolumn, superscriptaddress,"Cornell University, Ithaca, NY, 14853"]{revtex4}

\usepackage{amsmath}
\usepackage{empheq}
\usepackage[parfill]{parskip}
\usepackage{graphicx}
\usepackage[active]{srcltx}

\begin{document}
\newcommand{\bi}{\begin{itemize}}
\newcommand{\ei}{\end{itemize}}
\newcommand{\be}{\begin{equation}}
\newcommand{\fe}{\end{equation}}
\setlength{\parindent}{0.5cm}

\title{Synchronization as Aggregation: Cluster Kinetics of Pulse-Coupled Oscillators}

\author{Kevin P.  O'Keeffe}
\affiliation{Department of Physics, Cornell University, Ithaca, NY 14853, USA}

\author{Pavel L. Krapivsky}
\affiliation{Center for BioDynamics, Center for Polymer Studies, and Department of Physics, Boston University, Boston, MA 02215}

\author{Steven H. Strogatz}
\affiliation{Department of Mathematics, Cornell University, Ithaca, NY 14853, USA}

\date{\today}

\begin{abstract}
We consider models of identical pulse-coupled oscillators with global interactions. Previous work showed that under certain conditions such systems always end up in sync, but did not quantify how small clusters of synchronized oscillators progressively coalesce into larger ones. Using tools from the study of aggregation phenomena, we obtain exact results for the time-dependent distribution of cluster sizes as the system evolves from disorder to synchrony. 
\end{abstract}


\maketitle

In one of the first experiments on firefly synchronization, the biologists John and Elizabeth Buck captured hundreds of male fireflies along a tidal river near Bangkok and then released them at night, fifty at a time, in their darkened hotel room \cite{buck1968mechanism}. As they looked on in wonder, they observed that ``centers of synchrony began to build up slowly among the fireflies on the wall. In one area we would notice that a pair had begun to pulse in unison; in another part of the room a group of three would be flashing together, and so on.''  Synchronized groups continued to emerge and grow, until as many as a dozen fireflies were blinking on and off in concert. The Bucks realized that the fireflies were phase shifting each other with their flashes, driving themselves into sync.

Here we study stylized models of oscillators akin to the fireflies, in which synchrony builds up stepwise, in expanding clusters. By borrowing techniques used to analyze aggregation phenomena in polymer physics, materials science, and related subjects \cite{leyvraz2003scaling, krapivsky_book}, we give the first analytical description of how these synchronized clusters emerge, coalesce, and grow. We hasten to add, however, that the models we discuss are not even remotely realistic descriptions of fireflies; they are merely intended as tractable first steps toward understanding how clusters evolve en route to synchrony.  

Our work is part of a broader interdisciplinary effort \cite{pikovsky_book, strogatz_book}. Oscillators coupled by sudden pulses have been used to model sensor networks \cite{sensor_networks_time, low_powered_radios, firefly_sensor_networks, nishimura11, nishimura12}, earthquakes \cite{herz1995earthquake, bottani1997self}, economic booms and busts \cite{gualdi2014}, firing neurons \cite{timme09neurons, hopfield1995rapid}, and cardiac pacemaker cells \cite{peskin75}. Diverse forms of collective behavior can occur in these pulse-coupled systems, depending on how the oscillators are connected in space. Systems with local coupling often display waves \cite{bressloff97ring, bressloff1998waves}  or self-organized criticality \cite{corral95, bottani95, herz1995earthquake}, with possible relevance to neural computation \cite{hopfield1995rapid} and epilepsy \cite{worrell2002evidence}. In contrast, systems with global coupling, where every oscillator interacts equally with every other, tend to fall into perfect synchrony. Rigorous convergence results have been proven for this case \cite{mirollo90, gerstner93, ernst95, bottani95, bottani96}. But  the techniques used previously have not revealed much about the transient dynamics leading up to synchrony---the opening and middle game, as opposed to the end game. Aggregation theory offers a new set of tools to explore this prelude to synchrony.


We introduce a toy model, which we call \emph{scrambler oscillators}. It consists of $N \gg 1$ identical integrate-and-fire oscillators coupled all to all. Each oscillator has a voltage-like state variable $x$ that increases linearly according to $\dot{x} = 1$, rising from a baseline value of 0 to a threshold value of 1. Whenever any oscillator reaches threshold, it fires and does three things. (i) It kicks every other oscillator (and every cluster of oscillators) to a new random voltage, independently and uniformly---in this sense, it scrambles the other oscillators. (ii) It then ``absorbs'' any scrambled oscillators that lie within a distance $1/N$ of threshold, by bringing them to threshold and thereby synchronizing with them. To avoid the complications that would be caused by chain reactions of firings, we assume that the oscillators being brought to threshold do not get to fire until the \emph{next} time they reach threshold. (iii) The oscillator that fired resets to $x=0$ along with the oscillators it absorbed. 

If a cluster of $j$ oscillators does the firing, the same rules apply, except that now any oscillators within a distance $j/N$ of threshold get absorbed. The assumed proportionality to $j$ is natural, if each member of the cluster contributes to the pulse strength. We study other plausible coupling rules in the Supplemental Material~\cite{suppmat}.) 

The motivation for this scrambler model is that it leads to the simplest possible mean-field approximation. In the infinite-$N$ limit, we would like clusters of every size to be uniformly distributed in voltage at all times. This convenient property would greatly ease the derivation of the rate equations for the cluster kinetics. As we will see below, the predictions that follow from this approximation agree reasonably well with simulations.

Assume the initial voltages $x_i$, for $ i = 1, \dots, N$, are independent and uniformly distributed. At first, nothing interesting happens. The oscillators increase their voltages without interacting. But then one oscillator reaches threshold and fires. The remaining oscillators get scrambled, and perhaps some get absorbed. Then another oscillator fires, and so on. After this process has gone on for a while, the system has formed synchronized clusters of various sizes. 

Let $N_j(t)$ denote the number of clusters of size $j$ at time $t$. Thus there are $N_1(t)$ singleton oscillators, $N_2(t)$ pairs of synchronized oscillators, $N_3(t)$ triplets, and so on. The $N_j$ are correlated random quantities. They are correlated because oscillators belonging to clusters of one size are unavailable to clusters of another size, and they are random because of the randomness in the initial conditions and the scrambling procedure. It does not seem feasible to understand the  time-evolution of the $N_j$ unless they are so large that their fluctuations from one random realization to another are negligible. 

So assume from now on that $N_j\gg 1$ for all $j$ and replace these random quantities by their ensemble averages. Let $c_j = N^{-1}\langle N_j\rangle$ denote the average cluster densities.  One hopes that relative fluctuations are small; more precisely, $N^{-1} N_j = c_j +O(N^{-1/2})$. An even stronger assumption is that the densities of different sub-populations are asymptotically uncorrelated:
$N^{-2} N_iN_j = c_i c_j + O(N^{-1/2})$.

These $c_j$ allow us to define a natural disorder parameter, given by the total density $c(t) = \sum_j c_j(t)$. It measures the extent of the system's fragmentation. To see this, note that at $t = 0$ each oscillator is alone; only clusters of size 1 exist. Accordingly $c_1(0)=1$ and all other $c_j(0)=0$ for $j >1$. Hence $c(0)=1$, correctly indicating that the system starts out maximally fragmented. At the opposite extreme, as $t \rightarrow \infty $ only one giant cluster of synchronized oscillators exists. The system is then minimally fragmented: $c(t) = 1/N\rightarrow 0$ as $N \rightarrow \infty $.

To derive a rate equation for the decline of $c(t)$, let $R_i$ be the rate at which clusters of size $i$ fire, for $ i = 1, \dots, N$, and let  $L_i$ be the number of clusters lost to absorption in each such firing. Then $\dot{c} = - \sum_i R_i L_i$. 

To find $L_i$, recall that when a cluster of size $i$ fires, all the other clusters get assigned a new voltage uniformly at random. Moreover, any clusters assigned to the interval $[1-i/N, 1)$ get brought to threshold and absorbed. Since the voltages of these other clusters are uniformly distributed on $[0, 1]$, a fraction $i/N$ of them will be absorbed. There are $ \sum_j  N_j$ clusters in total. Hence the number absorbed is $L_i = (i / N) \sum_j  N_j = i \sum c_j = i c$. 

The rate $R_i$ takes more work to calculate. Since some clusters get absorbed, not every cluster gets the chance to fire. We must account for this depletion when calculating $R_i$. First consider the background rate of firing of clusters of size $i$ in the absence of absorptions. In other words, pretend for a moment that when an $i$-cluster fires, it simply scrambles every other cluster and restarts its own cycle without absorbing anyone. Call this background rate $R_i^0$. Since all oscillators move with velocity $v_i= \dot{x_i}=1$, and since the cluster density is $c_i$, the corresponding background rate of firing is $R_i^0 = c_i v_i = c_i$. Next, to find the actual $R_i$, we must subtract from $R_i^0$ the rate at which clusters of size $i$ are being absorbed and hence deprived of their chance at firing. Call this absorption rate $R_i^{a}$. Clusters of size $i$ are absorbed when clusters of size $j$ fire, for $ j = 1, \dots, N$, taking a fraction $j/N$ of the uniformly distributed $i$-clusters along with them. Since there are $N_i$ clusters of size $i$ and the $j$-clusters fire at rate $R_j$, the total rate at which $i$-clusters are being absorbed is given by $R_i^{a} = \sum_j ( j / N) N_i R_j = \sum_j j c_i R_j = c_i \sum j R_j$. 

Putting all this together gives $R_i = R_i^0 - R_i^{a} = c_i  - c_i \sum_j j R_j  = c_i(1 - \sum_j j R_j)$. Let $\beta = 1 -  \sum_j j R_j$. Note that $\beta$ is the same for all $i$, which enables it to be determined self-consistently, as follows.  From $R_i = \beta c_i$ we obtain $\beta = 1- \sum_j jR_j  = 1- \sum j( \beta c_j )$. Now invoke the identity $\sum_j j c_j = j (N_j/N) =  1$, which expresses conservation of oscillators. Solving for $\beta$ then gives $\beta = 1/2$ and therefore $R_i = c_i/2$.

\begin{figure}[h]
\centerline{\includegraphics[width=8cm]{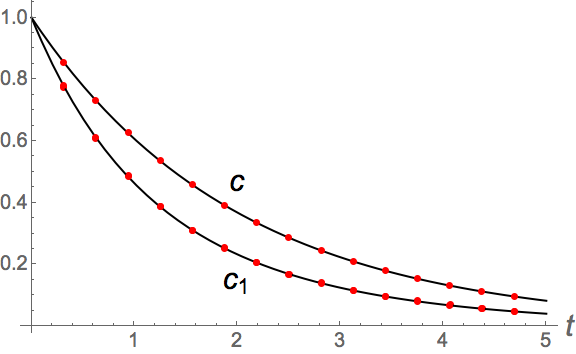}} 
\caption{\label{c_and_c1_plots} Theoretical and simulated $c(t)$ and $c_1(t)$. Solid lines show theoretical curves obtained analytically (see text). Data points show simulation results for $N = 10^4$ oscillators.}
\end{figure}

Next, plug the expressions derived for $R_i$ and $L_i$ into the rate equation $\dot{c} = - \sum_i R_i L_i$.  The result is $\dot{c} = - \sum_i (c_i / 2) (i c) =  -(c/2) \sum_i i c_i  = -c/2 $. Recalling that $c(0)=1$, we conclude that 
\be
c(t)= \exp(-t / 2 ).
\fe
Figure~\ref{c_and_c1_plots} shows this result matches simulations.

How do the individual cluster densities $c_i$  behave? To derive their rate equations, note that since the voltage space is the interval $[0,1]$, a segment of length $N^{-1}$ contains on average $N c \times N^{-1} = c$ clusters. In fact, the probability that it contains $n$ clusters (of any sizes) is given by the Poisson distribution: $\Pi_n = c^n e^{-c}/n!$. This is the mathematical expression of the assumption that clusters are distributed randomly without correlations. 

With this in mind, let us solve for $c_1(t)$, the density of singletons. It is the easiest $c_j(t)$ to analyze, since it can only decrease. Two mechanisms cause the loss of singletons: (i) a singleton fires and absorbs a cluster, and (ii) a cluster fires and absorbs $p \ge 1$ singletons. 

Consider mechanism (i). From the expression for $R_i$ obtained above, singletons fire at a rate $R_1 = c_1 /2$. When they fire, they absorb any cluster lying in the voltage segment $[1-1/N, 1)$. According to the Poisson distribution, the probability that this segment contains one or more clusters is given by $1- e^{-c}$.  Hence singletons are lost by mechanism (i) at a rate $(c_1/2) \left[1- e^{-c(t)} \right]$. 

The loss due to mechanism (ii) is handled similarly. Suppose $p$ singletons lie the interval $[1-j/N, 1)$ at the moment when a cluster of size $j$ fires, for $ j = 1, \dots, N$. This event happens with probability  $e^{-j c_1}(j c_1)^p / p!$, and when it does, it consumes $p$ singletons. The $j$-clusters fire at a rate $R_j = c_j/2$. Hence singletons are lost by mechanism (ii) at a rate
\be
\sum_{j \geq 1} \frac{c_j}{2} \times \sum_{p \geq 1} p \frac{(j c_1)^p e^{-j c_1}}{p!} = c_1 / 2.
\fe

Summing the loss rates from (i) and (ii) gives
 \be
 \frac{d c_1}{dt} = -\frac{c_1}{2}(2 - e^{-c(t)}).
 \fe
This equation has a closed-form solution in terms of exponential integrals:  
 \be
 c_1(t) = \exp(-t + \text{Ei}(-1) - \text{Ei}(- e^{-t/2})),
 \fe
where we have used the initial condition $c_1(0)=1$. Figure~1 shows good agreement between the theoretical and numerical $c_1(t)$.  

For $i>1$, the rate equation for $c_i$ includes gain terms as well as loss terms. Clusters of size $i>1$ can be created when two or more smaller clusters coalesce, or destroyed when they themselves coalesce with at least one other cluster. The loss term is a straightforward generalization of that for $c_1$, and is given by $(c_i/2) \left[2- e^{-i c(t)}\right]$.  
 
To find the gain term, imagine that a cluster of size $k$ fires. The segment $[1-k/N,1)$ may contain $a_1$ clusters of size 1, $a_2$ clusters of size 2, etc. This event happens with probability $\frac{(k c_1)^{a_1}}{a_1!} \times \frac{(k c_2)^{a_2}}{a_2!} \times \dots$ (where we are using the assumption that clusters of different sizes are independent as well as Poisson distributed). If the segment contains a combination of clusters such that $k + a_1 + 2a_2 + 3a_3  + \dots = i $, then a cluster of size $i$ will form. We sum over all such combinations for a cluster of size $k$ firing, and then sum over all $k$, to get the rate at which clusters of size $i$ are created:
\be
\sum_{k=1}^{i-1} \frac{c_k}{2} e^{-k c} \sum_{a_1 + 2 a_2 + \dots = i-k} \left( \prod_{p \geq 1} \frac{(k c_p)^{a_p}}{a_p!} \right).
\fe

\begin{figure}[h!]
  \centering
\includegraphics[width=8cm]{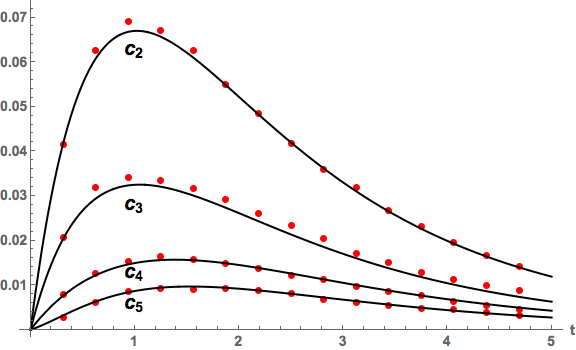}  
\caption{Theoretical and simulated cluster densities $c_2(t)$ through $c_5(t)$. Solid lines show theoretical predictions computed from numerical integration of Eq.~\eqref{ci_rate_eqns}. Data points show simulation results for $N = 5 \times 10^4$ oscillators.}
\label{higher_ci_plots}
\end{figure}

Combining the loss and gain terms, and transferring $c_i e^{-ic}$ into the gain term, we finally obtain
\be 
\dot{c_i} = - c_i + \sum_{k=1}^{i} \frac{c_k}{2} e^{-kc} \sum_{\sum p a_p = i-k} \left( \prod_{p \geq1} \frac{(k c_p)^{a_p}}{a_p !} \right).
\label{ci_rate_eqns}
\fe
We see from the sum that the equations \eqref{ci_rate_eqns} are recursive. They can be solved one by one, though not analytically, so we resort to numerical integration. Figure~\ref{higher_ci_plots} shows that the theoretical and simulated $c_i$ agree. 

Although we cannot find all the $c_i(t)$ explicitly, we can get their moments $M_n(t) = \sum_j j^n c_j(t)$. The first two moments are trivial: $M_0 = \sum_j c_j(t) = c(t)$ and $M_1(t) = \sum_j jc_j(t) = 1$, from conservation of oscillators. The higher moments can be obtained from a generating function. Let $G(z,t) = \sum_{k \geq 1} c_k(t) e^{k z}$. Then the infinite set of differential equations \eqref{ci_rate_eqns} transforms into 
\be
\frac{\partial G(z,t)}{ \partial t} +  G(z,t) = \frac{1}{2} G[z-c(t) + G(z,t), t].
\fe
This equation looks neat, but it is far from trivial, as the right-hand side involves $G$ in a very nonlinear manner. Using the identity $M_n(t) = \frac{\partial^n G}{\partial z^n} |_{z=0}$, we can however derive the following equations for the moments:
\be
\begin{split}
\dot{M_2} &= \frac{3}{2} M_2  \\
\dot{M_3} &= \frac{7}{2} M_3 + 3 (M_2)^2  \\
\dot{M_4} &= \frac{15}{4} M_4 +16 M_2 M_3 +\frac{3}{2} (M_2)^3.
\end{split} 
\label{moms}
\fe
Like the $c_i$ equations \eqref{ci_rate_eqns}, these moment equations are recursive and can be solved in succession. We find 
\be
\begin{split}
M_2(t) &= e^{3t/2}  \\
M_3(t) &= 7 e^{7t/2} - 6 e^{3t} \\
M_4(t) &= \frac{448}{5} e^{5t} - 128e^{9t / 2} + \frac{217}{5} e^{15t / 4} - 4 e^{27t/8}.
\end{split}
\fe

\begin{figure}[h!]
  \centering
    \includegraphics[width=8cm]{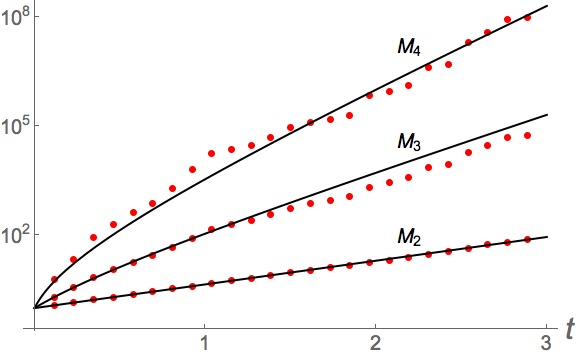}
      \caption{(Color online) Log plot of the first three nontrivial moments $M_2, M_3, M_4$. Black curves, theoretical predictions obtained from \eqref{moms}; red dots, average simulation results for 100 realizations of $N = 10^4$ oscillators.}
      \label{moments}
\end{figure}

Figure~\ref{moments} plots theoretical and simulated values of the $M_i$. The agreement is good for $M_2$ but worse for $M_3$ and $M_4$. This is to be expected. Each $c_i(t)$ is a stochastic process, subject to fluctuations dominated by the chance formation of big clusters. Since $M_n(t) = \sum_j j^n c_j (t)$, the higher moments amplify these fluctuations more and are therefore noisier themselves. 

Variants of the scrambler model can also be solved. For example, suppose that when a cluster of size $j$ fires, it absorbs all oscillators within a distance $k j/N$ of threshold, where $k>0$ is a tunable coupling strength. Or suppose that the pulse strength is independent of the size $j$ of the firing cluster. Both these cases are discussed in \cite{suppmat}. Nothing qualitatively new happens, but there are some interesting quantitative differences, e.g., in the latter case the disorder parameter $c(t)$ decays algebraically rather than exponentially. 

A more substantial modification would be to make the model deterministic. Scrambler oscillators respond randomly, but real oscillators respond deterministically, according to a phase-response curve \cite{winfree}. This deterministic character is incorporated in Refs.~\cite{peskin75, mirollo90} by assuming that when a cluster of size $j$ fires, it adds a voltage pulse $j \epsilon$ to every other oscillator, or pulls it up to threshold, whichever is less. For the case where $\epsilon = 1/N$ we show in \cite{suppmat} that this deterministic system can also be well approximated by the aggregation methods used above. The main new feature is that $c(t)$ and the cluster densities $c_i(t)$ become piecewise linear. But their overall shapes still resemble  those seen in the scrambler model.

There are many avenues to explore in future work. For example, throughout our analysis each oscillator obeyed $\dot{x_i} = 1$. Such linear sawtooth waveforms are reasonable for the oscillators used in sensor networks \cite{firefly_sensor_networks}, but not for neurons or cardiac pacemaker cells. Changing the concavity or shape of the waveform would make the analysis more difficult, and might even affect the cluster kinetics, because clusters would no longer be uniformly distributed as we have assumed throughout. Could different types of pulses be engineered to preserve a uniform distribution? Or perhaps different pulses would distribute the clusters in a nonuniform but still tractable manner. It would also be interesting to study cluster kinetics in oscillator systems with local coupling, network structure, heterogeneity, delays, and other realistic features.



\end{document}


\newcommand{\bi}{\begin{itemize}}
\newcommand{\ei}{\end{itemize}}
\newcommand{\be}{\begin{equation}}
\newcommand{\fe}{\end{equation}}
\setlength{\parindent}{0.5cm}

\title{Supplemental Material}

%
%
%
%
%


\maketitle
We study two modifications to the Scrambler oscillator model analyzed in the main text. The first is the original model, but with alternative couplings between oscillators. Specifically, we change how close to threshold an oscillator has to be in order to be absorbed by a firing cluster. This only modestly changes the analysis.

The second is a more substantial alteration. When an oscillator reaches threshold, it no longer scrambles every other oscillator to a new, random, voltage. Instead, it kicks every other oscillator up by a constant amount, or up to threshold, whichever is less. This deterministic resetting rule is in line with the simplest traditional models of pulse-coupled oscillators. As will be shown, this change makes for a more involved analysis.

\section{Alternative Couplings}
We first restate the dynamics of the original Scrambler model, and then describe the variations. 

Recall that in the main text we considered a population of $N \gg1$ integrate-and-fire oscillators coupled all-to-all. Each oscillator was characterized by a voltage-like state variable $x$, which increased linearly according to $\dot{x_i}=1$. When a cluster of $j$ oscillators reached a threshold value set to $1$, they fired and then instantly did three things: (i) they reassigned every other oscillator (or cluster of oscillators) a new voltage uniformly at random (they scrambled the oscillators) (ii) they absorbed any oscillators within a distance $j/N$ of threshold and (iii) they reset their voltage to 0 along with any oscillators they absorbed.

We now modify event (ii) in either of two ways: when a cluster of size $j$ fires, it either absorbs all oscillators within a new distance of threshold given by (a) $k j/N$ or (b) $k/N$. Modification (a) generalizes the original model by including an adjustable coupling $k$. Modification (b) assumes that the absorption region is independent of $j$, the size of the firing cluster. 

These generalizations change the analysis of the Scrambler model only slightly. For instance, to find the disorder parameter $c(t)$, we again use the rate equation $\dot{c} = - \sum_i R_i L_i$, where $R_i$ denotes the rate at which a cluster of size $i$ fires, and $L_i$ denotes the number of oscillators absorbed when a cluster of size $i$ fires. In the original model, to find $L_i$ and $R_i$, we made liberal use of the fact that all oscillators on the interval $[1- j/N, 1)$ were captured when a cluster of size $j$ fired. With the generalized couplings (absorption distances), this interval simply changes to $[1 - k j /N, 1)$ and $[1-k/N, 1)$. This change propagates through the analysis straightforwardly. Hence, we state the results in the following table without derivation. (The special function $W(x)$ in the table refers to Lambert's $W$ function.) \\

\begin{table}[h!]
\centering
\begin{tabular}{l*{6}{|c|}r}
             &$ j/N$ & $k j/N$ & $k/N$  \\
\hline
$L_i$            & $i c $ & $k i c$ & $ k c$   \\
$R_i $& $ c_i / 2 $ & $ c_i / (1+k) $ & $ c_i / (1 + k c)$    \\
$c(t)$           & $e^{-t/2}$ & $e^{- \frac{k}{ k+1} t}$ & $1 / (k W(k^{-1} e^{(k^{-1} + t) })) $  \\
\end{tabular}
\end{table}
\\ \\

\noindent
The rows of the table give the results for $L_i, R_i$ and $c(t)$; the columns show how the results vary for the three coupling schemes: original, (a), and (b). For coupling scheme (a), where the absorption distance is $kj/N$, the exponential decay constant in $c(t)$ is predicted to be $-k/(k+1)$. To test this, we simulated $c(t)$ for various $k$, and found the exponents of best fit. Figure 1 shows the results along with the theoretical curve. 

\begin{figure}[h!]
  \centering 
    \includegraphics[width=0.4\textwidth]{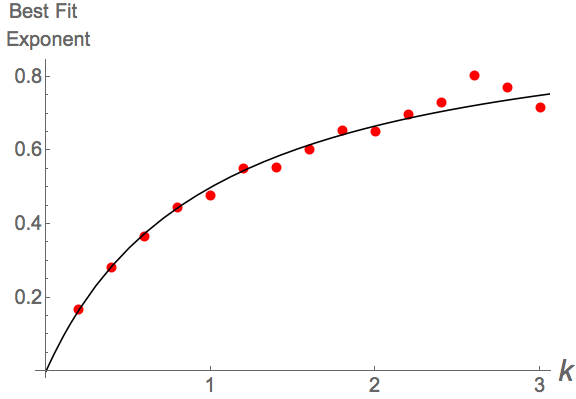}
     \caption{Magnitude of the decay constant in $c(t)$ for coupling scheme (a). Black curve, theoretical prediction $k/(k+1)$; red dots, simulation results for $N=5000$ oscillators.  } \label{mft}
\end{figure} 

\begin{figure}
  \centering 
    \includegraphics[width=0.4\textwidth]{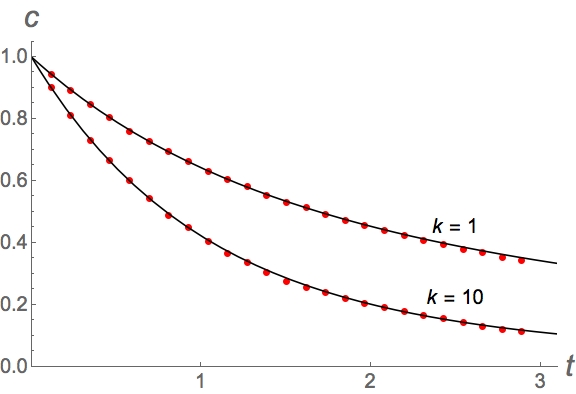}
     \caption{ Decay of the disorder parameter $c(t)$ for coupling scheme (b), which assumes a fixed absorption distance $k/N$. Black curves, theoretical predictions; red dots, simulation results  for $N=5000$ oscillators. Plots for $k = 1$ and $k = 10$ are shown. } \label{fixed_pulse}
\end{figure}

We find similarly good agreement between theory and simulations for coupling scheme (b), as shown in Figure~\ref{fixed_pulse}.  

Having solved for $c$, we could solve for the individual cluster densities $c_i$ and the moments $M_i$. Nothing qualitatively new happens (compared to what we saw in the main text for the original model), so we omit the details.

\section{Fictional Firefly Oscillators}
The Scrambler model is a toy model.  We introduced it to give the simplest possible mean-field model of pulse-coupled oscillators. Specifically, it was the random shuffling of oscillators during each firing event that simplified their analysis. It conveniently kept the voltages of all oscillators (and all clusters of oscillators) uniformly distributed on $[0,1]$ at all times.

The extreme randomness of this resetting rule, however, is contrived. Traditional models of pulse-coupled oscillators, such as those analyzed by Peskin~\cite{peskin75} and Mirollo and Strogatz~\cite{mirollo90}, obey deterministic resetting rules, much as real biological oscillators obey deterministic phase-response curves~\cite{winfree}. As we will show below, models with deterministic resetting can be reasonably approximated within the framework developed here. 

For simplicity, we will restrict attention to an almost absurdly idealized model of pulse-coupled oscillators, even more idealized than the models discussed in Refs.~\cite{peskin75, mirollo90}. It consists of what we will refer to as Firefly oscillators. (Fictional Firefly oscillators would be a more apt description, given that essentially everything about the model is unrealistic for real fireflies.) 

\subsection{Fireflies vs. Scramblers}
The equations of motion for the Fireflies are the same as for the Scramblers: $\dot{x_i} = 1$ (in between firing events). The initial voltages are again drawn from a uniform distribution. However, when a cluster of $j$ synchronized Fireflies reaches the threshold value of 1, that cluster does just two things: (i) it imparts a voltage pulse of size $j/N$ to all other oscillators. Any subsequent oscillators that reach threshold by virtue of this extra $j/N$, and hence get absorbed by the firing cluster, do not fire until the \textit{next} time they reach threshold. As before, this is to avoid complications that would be caused by chain reactions of firings.  (ii) The cluster resets to $x=0$ along with any oscillators it absorbed.

Will the Fireflies behave like the Scramblers? One difference between them has to do with how their speed $\dot{x}$ evolves as the system moves toward complete synchrony.  The average speed of the Fireflies doesn't remain constant at unity, as it does for the Scramblers. This is because when a cluster of size $j$ fires, all the other oscillators receive a pulse of size $j/N$ which boosts them up on their voltage curve (whereas the Scramblers were just randomly reassigned on $[0,1]$, and so keep the same speed $v_i=1$ on average). The speeds of the Scramblers are thus $v_i = 1 + v_{\mathrm{pulse}}$, where $v_{\mathrm{pulse}}$ must be determined. 

A second difference is that the firing rate of the Fireflies is piecewise constant (unlike that of the Scramblers, which as shown in the main text is a continuous function of time: $R_i(t)= c_i(t) / 2$.) To see why the firing rate for the Fireflies is piecewise constant, first define $\Delta x_i$ as the distance between the $i^{th}$ and $(i+1)^{th}$ oscillators (or cluster of synchronized oscillators). Second, divide the temporal evolution into periods $\{T_n\}$, where each period is the time taken for the full population to complete a full cycle. Then ask, how does $\Delta x_i$ behave during the first period, the time taken for the first wave of oscillators to complete their first cycle? Since all oscillators receive the same number of pulses and have the same speed, we see that each $\Delta x_i$ won't change while the oscillators complete their virgin ascent through [0,1]. This implies a constant firing rate during this first period. 

We will later show that this is not unique to the first period: the firing rates will take different, but constant, values during each period; they will be piecewise constant. This is in stark contrast to the Scramblers, where each $\Delta x_i$ is constantly changing as the oscillators get reshuffled on $[0,1]$ during each firing event. 

\subsection{Intuitive Picture of the Dynamics}
With these differences in mind, we begin with a qualitative description of the dynamics.  As mentioned, the speed of each oscillator, and the effect of a pulse on each oscillator, is the same. This means that the initial ordering of oscillators, or clusters of oscillators, will be invariant throughout the dynamics. They all march forward through $[0, 1]$ in a line with no passing. 

Then we consider the \emph{average} behavior as $N \rightarrow \infty$. In this limit, the average spacing between oscillators $\langle \Delta x_i \rangle $ approaches $ N^{-1}$. Now, what happens when the first oscillator fires? It captures all oscillators on the interval $[1-1/N, 1)$. Since $\langle \Delta x_i \rangle = N^{-1}$, there will be exactly one oscillator on this interval. So, on average, one oscillator will be captured. This procedure will repeat itself for the next oscillator that fires, and the oscillator after that, such that every oscillator that fires captures the one behind itself. In this mean-field sense, then, the first wave of oscillators will be an orderly sequence of fire/capture/fire/capture, so that at the end of the first period, all oscillators will have synchronized into pairs spaced equally apart. 

Of course, $\Delta x_i$ will have fluctuations about the mean value of $N^{-1}$. For the oscillators spaced such that $\Delta x_i < N^{-1}$, no captures will take place. For $\Delta x_i > N^{-1}$, at least one capture will take place, and possibly more. So, at $t = T_1$  there will be a number of clusters of different sizes. It is not clear how these clusters are distributed on $[0, 1]$. Say, for example, that mostly clusters of size $2$ and $3$ formed, while a cluster of size 4 was the first capture, and a cluster of size 6 was the last capture. Then the clusters of sizes 2 and 3 will be approximately uniformly distributed in voltage, while the distribution of those of sizes 4 and 6 will be more sharply peaked.

Nevertheless we assume that the clusters of size $i$ are uniformly distributed on $[0,1]$ for each $i$. We recognize that this won't be accurate for each $i$ for all values of $t$. It will however be accurate for those values of $i$ which contain most of the oscillator ``mass'' and less so for those with less of the mass. So, our assumption will be imprecise for those $c_i$ which are small, but since they are small, the inaccuracy won't matter, to first order. 

Now that we understand the first period, how will the second period proceed? We again consider the \emph{average} behavior as $N \rightarrow \infty$. In this mean-field description, we earlier concluded that at the end of the first period, all oscillators would have synchronized into pairs spaced equally apart on $[0,1]$. When the first pair fires, therefore, there will again be exactly one pair of synchronized oscillators behind them, and so as before, the second period will be an orderly sequence of fire/capture/fire/capture, resulting in all oscillators having synchronized into clusters of size 4. 

Continuing this logic, we see the size of clusters will double during each period. Moreover, we observed that the clusters of oscillators will begin each period evenly spaced from each other. This implies the aforementioned piecewise constant firing rate. The mean-field dynamics are therefore trivial: there is a train-like progression of clusters of the same size through $[0,1]$, with each cluster that reaches threshold doubling in size by absorbing the cluster behind it.

\subsection{Mean-Field Analysis}
With this picture in mind, we analyze the rate equation for our disorder parameter: $\dot{c} = -\sum_i R_i L_i$. 

To calculate $L_i$, we again find all oscillators in the interval $[1-i/N, 1)$. It is tempting to write down $L_i(t) = \sum_j (i/N) N_j(t)$. This isn't strictly true however, because as we observed, $\Delta x_i$ will be constant during each period, which means we must evaluate $N_j$ at the start of said period. To make this clear, define $\tilde{x}(t) = x(t = T_{n-1})$ for $T_{n-1} < t < T_n$. The tilde notation signifies that throughout the $n^{th}$ period the quantity $x$ is fixed at its value at the start of that period. In terms of this tilde notation, the desired result is $L_i = \sum_j (i/N) \tilde{N_j}(t) = i \tilde{c}$.

To find the firing rate, we follow the procedure used in the main text for the Scramblers: we decompose the firing rate into two parts: $R_i = R_i^0 - R_i^{a}$. The rate of firing in the absence of absorption will be $R_i^0 = \tilde{c_i} v_i = \tilde{c_i}(1+v_{\mathrm{pulse}})$. The absorption rate will again be $R_i^{a} = \sum_j (j/N) \tilde{N_i} R_j = \tilde{c_i} \sum_j j R_j$. 

We next determine $v_{\mathrm{pulse}}$. The ``pulse velocity'' due to a cluster of size $j$ will be (absolute number of pulses per sec}) $\times$ (distance per pulse). Since $R_j$ is the rate of firing of $c_i$,  $R_j N $ will give the absolute number of fires. The distance per pulse is $j/N$. The total pulse velocity is thus $v_{\mathrm{pulse}} = \sum_j (N R_j) (j/N) = \sum_j jR_j$. This gives $R_i^0 = \tilde{c_i}(1 +  \sum_j j R_j) $. Putting all this together, we find
\be
R_i = \tilde{c_i}(1 +  \sum_j j R_j) - \tilde{c_i} \sum_j j R_j \\
\fe 
which reduces to $R_i = \tilde{c_i}$.

Substituting our expressions for $R_i$ and $L_i$ into the rate equation for $c$ then yields
\be \label{c}
\dot{c} &= - \sum_i L_i R_i = - \sum_i i \tilde{c} \tilde{c_i}  = - \tilde{c} \sum_i i \tilde{c_i}= - \tilde{c} \\ 
\fe
and hence $\dot{c} = - \tilde{c}.$
Thus, we see that $\dot{c}$ will be a piecewise linear function. 

To solve for this function, we need to determine the periods $\{T_n\}$. The average speed is  $v = 1 + v_{\mathrm{pulse}} = 1 + \sum_j j R_j =1 + \sum_j j c_j = 2$. This gives $\{T_n\} = \{0, 0.5, 1, \dots\}$. For these values of $\{T_n\}$ and the initial condition $c(0)=1$, the solution of Eq.~\eqref{c} is the piecewise linear function


\be \label{ff_c}
c(t) &= \frac{p+2-2t}{2^{p+1}}, \; \; \; \; \; \frac{p}{2} < t < \frac{p+1}{2} 
\fe
where $p = 0, 1, 2, \dots$. Hence, for the Firefly model, the disorder parameter $c(t)$ is a series of line segments of length $0.5$, with slopes that are sequentially reduced by a factor of 2.

\begin{figure} [h!]
  \centering 
    \includegraphics[width=0.4\textwidth]{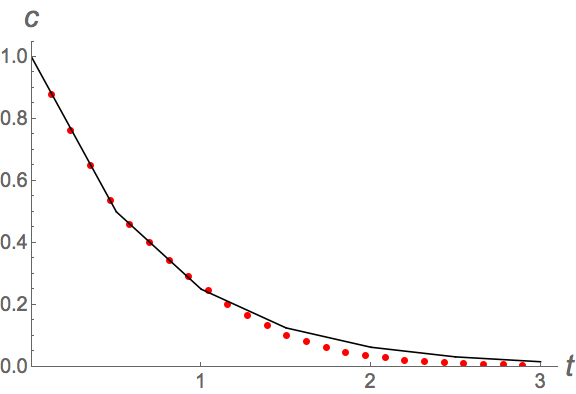}
     \caption{ Piecewise linear behavior of the disorder parameter $c(t)$ for the Firefly model. Black curve, theoretical prediction~\eqref{ff_c};  red dots,  simulation results for $N = 5000$  oscillators.} \label{mft}
\end{figure}

We stress however that our results for $L_i, R_i, v_{\mathrm{pulse}}$, and $c(t)$ are only leading-order approximations, based on mean-field arguments. We expect fluctuations around these values. Our hope is that these will be small. 

Figure~\ref{mft} shows the simulated behavior of $c(t)$ against the mean-field prediction~\eqref{ff_c}. As can be seen, there is reasonable agreement until late times.





Now that we have $c(t)$, the next target is $c_i(t)$. Carrying out the same analysis as for the Scramblers, we find 
\be \label{ff_ci}
 \dot{c_i} &= - 2\tilde{c_i} + \sum_{k=1}^{i} \tilde{c_k} \; e^{-k \tilde{c}} \sum_{\sum p a_p = i-k} \left( \prod_{p \geq1} \frac{(k \tilde{c_p})^{a_p}}{a_p !} \right ).\\
\fe
Since the quantities on the right hand side are held fixed over each period, solving for each $c_i$ is straightforward. Figure~\ref{ff_cis} shows the resulting solutions along with simulated values. Reasonable agreement is evident. We restate that our results are mean-field equations, so some discrepancy is expected.
The moments $M_i(t)$ will also be piecewise linear, and can be obtained in a similarly straightforward manner, following the methods shown in the main text. 

\begin{figure} \label{ff_cis}
  \centering 
    \includegraphics[width=0.45\textwidth]{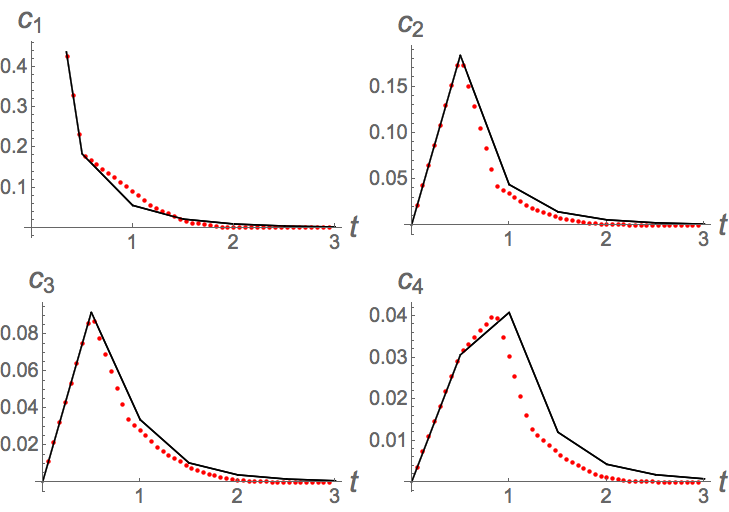}
\caption{ Time evolution of the cluster densities $c_1$ through $c_4$ for the Firefly model with uniformly random initial conditions. Black curves, theoretical predictions derived from Eq.~\eqref{ff_ci}; red dots; simulation results for for $N = 3 \times 10^4$ oscillators. The discrepancies are due to finite-$N$ effects.}
\label{ff_cis}
\end{figure}

